\documentclass[12pt,preprint]{aastex}

\usepackage{CJK}

\usepackage{epsfig}
\usepackage{amsmath}

\shorttitle{$\kappa$ electrons in PNe}
\shortauthors{Zhang et al.}

\begin{document}

\title{
On the non-thermal $\kappa$-distributed electrons in planetary nebulae and
\ion{H}{2} regions: the $\kappa$ index and its correlations with other nebular properties
}

\begin{CJK*}{UTF8}{gbsn}

\author{Yong Zhang (张泳)$^{1,2}$ Bing Zhang (张兵)$^{3}$, and Xiao-Wei Liu (刘晓为)$^{3,4}$}
  \altaffiltext{1}{Space Astronomy Laboratory, Faculty of Science, The University of Hong Kong, Pokfulam Road, Hong Kong, China; zhangy96@hku.hk} \altaffiltext{2}{Department of Physics, The University of Hong Kong, Pokfulam Road, Hong Kong, China}
\altaffiltext{3}{Department of Astronomy, Peking University, Beijing 100871, China}
\altaffiltext{4}{Kavli Institute for Astronomy and Astrophysics, Peking University, Beijing 100871, China}

\begin{abstract}

Recently, a suspicion arose that the free electrons in planetary 
nebulae (PNe) and \ion{H}{2} regions might have non-thermal energy 
distributions. In this scenario, a $\kappa$ index is introduced to
characterize the electron energy distributions, with smaller $\kappa$ 
values indicating larger deviations from Maxwell-Boltzmann distributions.
Assuming that this is the case, we determine the $\kappa$ values for a sample 
of PNe and \ion{H}{2} regions by comparing the intensities of [\ion{O}{3}] collisionally 
excited lines and the hydrogen Balmer jump. We find the average $\kappa$ indices of
PNe and \ion{H}{2} regions to be 27 and 32, respectively. Correlations between the resultant $\kappa$ values and various physical properties 
of the nebulae are examined to explore the potential origin of non-thermal electrons in
photoionized gaseous nebulae.  However, no positive result is obtained.
Thus the current analysis does not lend
 to support to the idea that $\kappa$-distributed
 electrons are present in PNe and \ion{H}{2} regions.

\end{abstract}


\maketitle
\end{CJK*}

\section{Introduction}

The accurate determination of element abundance in planetary nebulae (PNe) and 
\ion{H}{2} regions is important for understanding the matter cycle between stars and the ISM. This crucially depends on the assumed free electron energy distributions (EEDs) in that the emissivities of recombination lines (RLs) and collisionally excited lines (CELs) are given by the integral of the known energy dependence of the relevant atomic cross section over the EEDs. In the past, the generally accepted assumption is that free electrons in photoionized nebulae are in thermal equilibrium, in which the EEDs follow a Maxwell-Boltzmann (MB) 
distribution, characterized by a single parameter, the electron temperature.
However, this assumption has never been validated by direct observations, but is largely based on the theoretical argument that the electron thermalization 
time-scale via elastic collisions is much shorter than the other known 
processes that produce non-thermal electrons under typical nebular conditions.
Although there were some early debates about the MB distributions in PNe \citep[see Introduction of][]{ss14}, it has not been seriously doubted until recently.

It has been long known that there exist non-MB distributed particles in solar and space plasmas
\citep[e.g.][]{va68,fa79,sf87,cr14}. The $\kappa$ distributions, consisting of a low-energy MB core and 
a high-energy power law tail,
are introduced to characterize those non-MB distributions,
with smaller $\kappa$ indices indicating larger deviations from MB
distributions. \citet{leu02} and \citet{liv09} show that the $\kappa$ 
distributions can arise naturally  from Tsallis' nonextensive statistical 
mechanics. In non-isolated systems, on-going accelerations, such 
as those induced by magnetic reconnections, shocks, and turbulence, can generate non-thermal
electrons, which are predicted to have $\kappa$ EEDs according to the 
non-equilibrium generalization of the MB distributions. This
raises a critical question of whether the $\kappa$ distributions also
widely exist in other astronomical environments other than the solar
system plasmas, such as photoionized gaseous nebulae, and the MB distributions
are only a special case of equilibrium.

Inspired by the studies of solar system plasmas, 
\citet{nic12,nic13} proposed that the $\kappa$ distributions provide a 
potential solution for 
the long-standing problem in nebular physics, viz., 
the discrepancies of electron temperatures and heavy-element abundances derived from RLs and CELs. 
The RL/CEL discrepancy is commonly quantified by the ratio of 
O$^{2+}$ abundances obtained from RLs and CELs, called abundance discrepancy
factor \citep[ADF;][]{liu06}. The ADFs range from 1 to 71, with the most extreme case found in the PN
Hf\,2-2 \citep{lb06}. As the high-energy tail of $\kappa$ EEDs preferably contribute 
to the emission of CELs relative to RLs, CEL diagnostics will result in an overestimated
electron temperature, leading to underestimated element abundances,
when ones use MB distributions to interpret the spectra actually produced by 
$\kappa$ distributions. Later on, $\kappa$ distributions were invoked to
explain the abundance discrepancies in \ion{H}{2} regions derived from
strong-line and $T_{\rm e}$-based methods \citep{ds13}, and the abnormal 
\ion{C}{2} line ratios in Type 2 quasars \citep{hb14}. The $\kappa$ indices represent 
an additional free parameter to match model to observations. In order to examine
the validation of this solution, it is important to develop methods to diagnose
the EEDs in photoionized gaseous nebulae.

The first attempt to discriminate EEDs in PNe was made by \citet{ss13}, who investigated 
the intensities of \ion{C}{2} dielectronic recombination lines under $\kappa$
and MB distributions and compared the theoretical predictions with the spectra of a
few PNe. However, taking into account the uncertainties, it is hard to 
distinguish the two scenarios from the limited observations. Therefore, they
cannot draw any definite conclusion.

Another approach to sample the EEDs in PNe is to simulate the \ion{H}{1} free-bound (FB) 
emission \citep[][hereafter ZLZ]{zl14}, i.e. the nebular optical continua. 
Using that method, ZLZ found that the observed \ion{H}{1} FB spectra of four 
PNe with extremely large ADFs cannot be interpreted in terms of a single MB 
EED, but are equally well fitted by either a
two-component MB model or by $\kappa$-distributed electrons. However, 
shortly thereafter, a similar study independently undertook by \citet[][hereafter SS]{ss14} 
shows that the two-component MB model actually better accounts for
the \ion{H}{1} FB  spectrum of Hf\,2-2 than $\kappa$-distributed 
electrons, in contrast to ZLZ's conclusion. \citet{ss15a} attributed the 
discrepancy to the different treatments of H11 emission coefficient,
 $\alpha_{\rm eff}$(H11).
In both models of ZLZ and SS, the continuum intensity has been 
normalized to the integrated intensity of the H11 Balmer line at 
3770\,{\AA}. Following \citet{nic12}, ZLZ used an approximate 
function to obtain $\alpha_{\rm eff}$(H11) for a $\kappa$ distribution.
The approximation is generally sufficiently accurate, but
at high temperatures and/or extremely low $\kappa$ indices a correction factor
should be considered (see Table~1 of SS). At the typical nebular temperature of $10^4$\,K,
the correction increases rapidly from  2 to 30 percent as the $\kappa$ index 
decreases from 10 to 2. However, we do not  think that this is the main
reason causing the inconsistent results between ZLZ and SS.
In the first place, it seems unlikely (although not impossible) that
the spectrum fitting based on $\kappa$ distribution can be greatly
improved by using a more inaccurate $\alpha_{\rm eff}$(H11). Moreover,
the model fitting largely relies on the slope of \ion{H}{1} FB continuum,
and the approximate treatment of H11 only results in an uncertain
scaling constant and thus only has minor effect on the result. We note that
the primary difference between the modelling approaches of
ZLZ and SS is in the treatment of subtraction of the underlying stellar light
contaimination. The spectral energy distribution of 
contaminating stellar continuum has
been assumed to follow a power-law, $I_{\lambda,\rm star}\sim\lambda^{-\beta}$.
SS  obtained $I_{\lambda,\rm star}$ by fitting the model continuum to the 
observed one in two wavelength segments longer than the Balmer edge  
(3780--3790 and 4180--4230\,{\AA}).  For this purpose, they need to 
presuppose a 
\ion{H}{1} FB continuum for a given temperature and $\kappa$ value.
In contrast, ZLZ considered the whole line-free spectral range 
3550--4200\,{\AA},
and optimized the model fitting by adjusting the ${\beta}$ value.
The latter may provide a more flexible way to fit the global 
spectral behavior.  It seems to us this different treatment of the underlying
stellar light contamination 
 might actually be the primary cause for the different conclusions drawn by
SS and ZLZ. We should point out that accurate subtraction of background 
contribution and reliable continuum flux calibration are the main challenge 
of using \ion{H}{1} FB continuum  to probe the EEDs.

Although no solid observational evidence is found to support the idea of
deviation from MB EEDs in nebulae, if the $\kappa$ distributions were proven valid,
one would need a new paradigm for the study of nebular astrophysics. In this
paper, we present  determinations of  the $\kappa$ indices for a sample
of PNe and \ion{H}{2} regions under the assumption that the $\kappa$
EEDs are fully responsible for the RL/CEL discrepancy problem. The reported
$\kappa$ values could serve as a basis for future studies of this problem.
The rest of the paper is organized as follows. In Section~2 we describe the
methodology to derive the $\kappa$ indices from nebular spectra.
The results are reported in Section~3. In Section~4 we explore
the correlations of the resultant $\kappa$ values and other nebular
properties, in order to investigate the potential causes of 
$\kappa$ EEDs.  A summary is given in Section~5.

\section{METHODOLOGY}

\subsection{The $\kappa$ distribution and its effects on plasma diagnostics}

As quoted, the calculations of RL and CEL intensities largely depend on 
EEDs. The EED is quantified by $f(E)$, the fraction of electrons having 
an energy of $E$.  In an equilibrium state, $f(E)$ is given by the MB 
function 
\begin{equation}\label{MB1}
f_{\rm MB}(E)=2\sqrt{\frac{E}{\pi}}\left(\frac{1}{k_BT_{\rm e}}\right)^{3/2}\exp\left(\frac{-E}{k_BT_{\rm e}}\right),
\end{equation}
where $k_B$ is the Boltzmann constant, and $T_{\rm e}$ is the electron temperature.  
When the plasma is in a non-equilibrium stationary state, the EED follows a
$\kappa$ distribution defined by
\begin{equation}\label{k1}
f_\kappa(E)=2\sqrt{\frac{E}{\pi}}\left(\frac{1}{k_BT_U}\right)^{3/2}
\frac{\Gamma(\kappa+1)}{(\kappa-3/2)^{3/2}\Gamma(\kappa-1/2)}\left[1+\frac{E}{(\kappa-3/2)k_BT_U}\right]^{-\kappa-1},
\end{equation}
where $\Gamma$ is the gamma function, $\kappa$ is a parameter $>1.5$
describing the degree of departure from the MB distribution, and 
$T_{\rm U}$ is the non-equilibrium temperature characterizing the
mean kinetic energy. Lower $\kappa$ values represent larger departures
from the MB distribution. As the $\kappa$  index approaches infinity,
the $\kappa$ distribution decays to the MB distribution, where we have
$f_\kappa(E)=f_{\rm MB}(E)$ and $T_{\rm U}=T_{\rm e}$.
 
Compared to the MB distribution, the $\kappa$ distribution has a high-energy tail that 
is power-law distributed. For a given $T_{\rm U}$ value, the fraction of high-energy
electrons increases with decreasing $\kappa$. The low-energy parts of a $\kappa$ distribution can be 
approximated by a MB distribution of a temperature $T_{\rm core}=(1-1.5/\kappa)T_{\rm U}$. Since
$\int f_\kappa(E)dE \equiv f_{\rm MB}(E)dE \equiv 1$, this
MB function should be scaled by a factor of $R<1$ to match the core of the $\kappa$ distribution, where
$R$  can be obtained by requiring $Rf_{\rm MB}(k_BT_{\rm core})=f_\kappa(k_BT_{\rm U})$. From
Equations~(\ref{MB1}) and (\ref{k1}), we have
\begin{equation} \label{eq3}
R=2.718\frac{\Gamma(\kappa+1)}{\kappa^{3/2}\Gamma(\kappa-1/2)}\left(1+\frac{1}{\kappa}\right)^{-\kappa-1}.
\end{equation}
The $(1-R)$ value represents the fraction of non-thermal electrons, which ranges from 0.01 to 0.45 when 
the $\kappa$ index decreases from 100 to 2.

It is conceivable that incorrect results will be  obtained if ones utilize the MB-based method to determine 
the electron density ($N_{\rm e}$) and $T_{\rm e}$ of the plasma with the $\kappa$ EED\footnote{Hereafter, for simplicity, unless otherwise specified, the plasmas studied in this paper refer to those with $\kappa$ EEDs.}.
If the $N_{\rm e}$-diagnostic lines are dominantly excited by low-energy electrons
(e.g. RLs and infrared fine-structure CELs), the MB-based method will result in a $N_{\rm e}$ value that 
is overestimated by a factor of $1/R$. The Balmer jump (BJ) of \ion{H}{1} recombination spectrum
and the [\ion{O}{3}] nebular and auroral lines are the most used $T_{\rm e}$-diagnostics in nebulae.
The former is mostly contributed by low-energy electrons, and thus mainly measures the low-temperature MB core,
while the latter arises from collisional excitation of high-energy electrons.
It follows that ones may obtain a lower $T_{\rm e}$(BJ) relative to  
$T_{\rm e}$([\ion{O}{3}]) under the traditional assumption of MB EEDs. 


\subsection{Determination of $\kappa$ indices}

The $\kappa$ and $T_{\rm U}$ values of an isotropic plasma can be derived by comparing
the BJ intensity, $J_{\rm B}= [I(\lambda3650^-)-I(\lambda3650^+)]/I({\rm H11})$\,{\AA$^{-1}$},
and the nebular-to-auroral line ratio of [\ion{O}{3}], $R$([\ion{O}{3}])$=$[\ion{O}{3}]$\lambda\lambda(4959+5007)/4363$. 
\citet{zhang04} have simulated the \ion{H}{1} FB spectrum with MB EEDs. A similar method can
be used to calculate the BJ with $\kappa$ EEDs, in which $f_{\kappa}(E)$  was substituted for $f_{\rm MB}(E)$.
The dependence of the BJ intensity on $\kappa$ and $T_{\rm U}$ has been elucidated in ZLZ.
At a particular $T_{\rm U}$  value, a lower $\kappa$ index leads to a higher $J_{\rm B}$.
We use the same approach as in ZLZ to simulate the BJ intensity. In order to normalize $J_{\rm B}$, we take
the effective recombination coefficients of H11 under $\kappa$ EEDs, $\alpha^{\rm eff}_\kappa$(H11), 
recently tabulated by \citet{ss15a}. For either recombination or collision processes, the rate coefficient is 
given by
\begin{equation}\label{eq4}
\alpha_\kappa=\int \sigma(E)\sqrt{\frac{2E}{m_{\rm e}}}f_\kappa(E)dE,
\end{equation}
where $m_{\rm e}$ is the electron mass, and $\sigma(E)$ is the relevant cross-section. 
In order to simulate the \ion{H}{1} FB transition, the recombination cross-section, $\sigma(E)$,
is determined from the photoionization  cross-section through the Milne relation.
Then $\alpha^{\rm eff}_\kappa$(H11) can be deduced by solving the collisional-radiative recombination problem.
\citet{ss15a} performed such calculations, and derived the $\alpha^{\rm eff}_\kappa$(H11) values at various
combinations of $\kappa$, temperature, and density.  
The dependence of $R$([\ion{O}{3}]) on $\kappa$ and $T_{\rm U}$ can be obtained by solving
the level populations for the five-level atomic model,  
for which we require the effective collision strengths with $\kappa$ EEDs. In principle,
the effective collision strengths can be determined from Equation~(\ref{eq4})
if we have the collisional cross-section, $\sigma(E)$.  However, the cross sections are 
difficult to be tabulated and are not always available in literature. Recently, \citet{ss15b} 
calculated the $\kappa$-dependent effective collision strengths for electron collision 
excitation of the [\ion{O}{3}] CELs. We have taken their reported data for the calculations
of $R$([\ion{O}{3}]). Since the non-thermal electrons would strengthen the auroral lines
more than the nebular lines, the $R$([\ion{O}{3}]) value decreases with decreasing
$\kappa$ values.

Figure~\ref{fig1} plots the theoretical predictions of $J_{\rm B}$ versus  $R$([\ion{O}{3}]) as  functions of $\kappa$ and $T_{\rm U}$, as well as those of MB EEDs. 
The calculations are  essentially independent of $N_{\rm e}$ because the 
 critical densities for the involved lines are much larger than the typical nebular density.
Therefore, we have assumed a constant $N_{\rm e}$ of 10$^3$\,cm$^{-3}$ for all the calculations.
An inspection of Figure~\ref{fig1} shows that $J_{\rm B}$ and $R$([\ion{O}{3}]) are more sensitive to the $\kappa$ index at low temperature.
With increasing $T_{\rm U}$, the peak of EEDs shift towards higher
energy, and thus more electrons with energy lying within the MB core
can contribute to the excitation of [\ion{O}{3}] CELs. As a result, the 
$J_{\rm B}$ versus  $R$([\ion{O}{3}]) relations would converge to that of MB EEDs
at higher temperature. As shown in Figure~\ref{fig1}, the  $\kappa$ index can be determined 
with a reasonable degree of accuracy from the observed $J_{\rm B}$ and $R$([\ion{O}{3}])
in the regions of $\kappa<60$ and $T_{\rm U}<15000$\,K.

The basic idea of this method is similar to that of \citet{nic12} through
comparing  $T_{\rm e}$(BJ) and $T_{\rm e}$([\ion{O}{3}]). But the approach
of \citet{nic12} contains some approximations, e.g., the rate coefficients
are approximately obtained (see Section~1 and the discussion in SS), and they 
have assumed $T_{\rm e}({\rm BJ}) = T_{\rm core}$. These approximations are
generally justified, but would lead to larger uncertainties for low $\kappa$ index.
In contrast, we have used the most recently reported atomic data, and thus are 
able to provide a more robust estimate of the  $\kappa$ index.

Based upon the $T_{\rm e}$(BJ) versus $T_{\rm e}$([\ion{O}{3}]) diagrams, 
\citet{nic12}  concluded that PNe and \ion{H}{2} regions have typical $\kappa$ indices 
of larger than 10. However, they did not present the $\kappa$ values for individual 
nebulae.  In the present work we attempt to determine the $\kappa$ indices for a sample
of gaseous nebulae, including 82 PNe and 10 \ion{H}{2} regions. The data 
utilized were taken from work recently published by our research group and others
(see Tables~\ref{PN} \& \ref{H2}). These high signal-to-noise spectra have been 
obtained with the purpose of investigating the RL/CEL discrepancy problem.
With careful flux calibrations and dereddening corrections as well as
accurate measurements of the Balmer Jump, they are particularly well suited
for our analysis. In some cases, where the $J_{\rm B}$ values are not explicitly given
in the literature, we deduced the $J_{\rm B}$ values through the
$T_{\rm e}$(BJ) versus $J_{\rm B}$ relation of \citet[][see their Equation~(3)]{ll01}.

\section{RESULTS}

\begin{figure}
\plotone{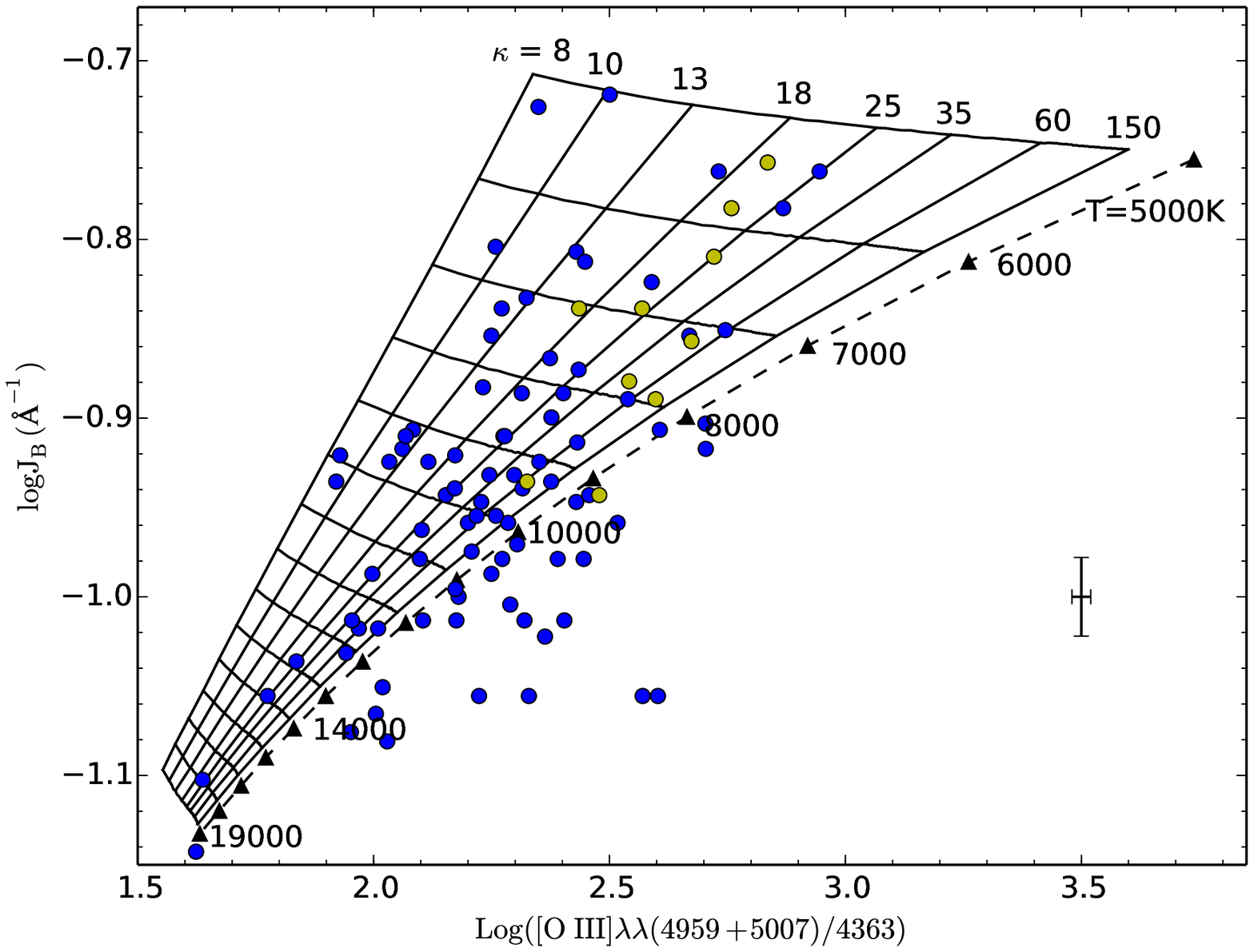}
\caption{The intensities of the Balmer Jump versus the [\ion{O}{3}] line ratios
as functions of $\kappa$ and $T_{\rm U}$. The dashed curve with filled triangles
shows the theoretical predictions from MB electron distributions at different
temperatures. The filled circles represent the observations of  PNe (blue) and \ion{H}{2} 
regions (yellow).  A typical error bar is shown in the lower right corner.
\label{fig1}}
\end{figure}

The observed $J_{\rm B}$ and $R$([\ion{O}{3}]) values are overplotted in Figure~\ref{fig1}.
As the data are taken from different references, the measurement uncertainties are not
available for all the observed values. 
We roughly estimate a typical error  bar from our spectra,
 as shown in the lower right corner of this figure. The  [\ion{O}{3}] CELs are
relatively strong, and thus $R$([\ion{O}{3}]) can be obtained with high accuracy.
The uncertainty of the $\kappa$ index mainly comes from the measurement of the
Balmer Jump, and it steeply increases with increasing $\kappa$ and $T_{\rm U}$.
If the EEDs follow single MB distributions,  all the plotted points should lie on
the dashed curve in Figure~\ref{fig1}. However, most of them fall on the upper
left of that curve, which can be explained in terms of $\kappa$ EEDs. There are a very small 
number of data points lying on the far lower right of the dashed curve, e.g. the regions of
$T_{\rm e}({\rm BJ})>T_{\rm e}$([\ion{O}{3}]). We can tentatively attribute this to
shock heating in the outer low-ionized regions and/or the possible existence of metal-poor clumps,
although the exact reason remains unclear.

The resultant $\kappa$ and $T_{\rm U}$, along with the other nebular properties, are 
given in Tables~\ref{PN} \& \ref{H2} for PNe and \ion{H}{2} regions, respectively.
In these tables, $N_{\rm e}$ is the average value of electron densities obtained by 
various $N_{\rm e}$-diagnostics available in the literature, and the effective
density  under $\kappa$ EEDs, $N_{\rm eff}$, is obtained through multiplying $N_{\rm e}$ by a factor of 
$R$ (see Section~2.1).
In Figure~\ref{fig1}, some  data points are too close to the MB predictions to allow
reliable determinations of $\kappa$, and we can only estimate a lower limit of 60. Excluding
those with $\kappa>60$, we finally give the $\kappa$ indices of 47 PNe and 8 \ion{H}{2} regions.
The errors of $\kappa$ can be inferred graphically from Figure~\ref{fig1}.  
It should be cautious in using the resultant values in the high-$\kappa$ and/or high-$T_{\rm U}$ regions
for further analysis.
As is clear in Figure~\ref{fig1}, PNe distribute in  more scattering and higher
temperature regions than \ion{H}{2} regions in the ($\kappa$, $T_{\rm U}$) parameter space.
We obtain the average values of $\kappa=27$ and $32$ for the 47 PNe and  
8 \ion{H}{2} 
regions, respectively, suggesting that PNe probably have a systematically greater departure
from MB EEDs than \ion{H}{2} regions. We did not find very extreme PNe with $\kappa<8$.
The average $\kappa$ index in nebulae is far larger than those in most of the other space plasmas
\citep[see Table~1 of][]{liv15}, but is close to that of the low solar corona
\citep[10--25;][]{cr14}.

%
%
%
%

\begin{figure}
\plotone{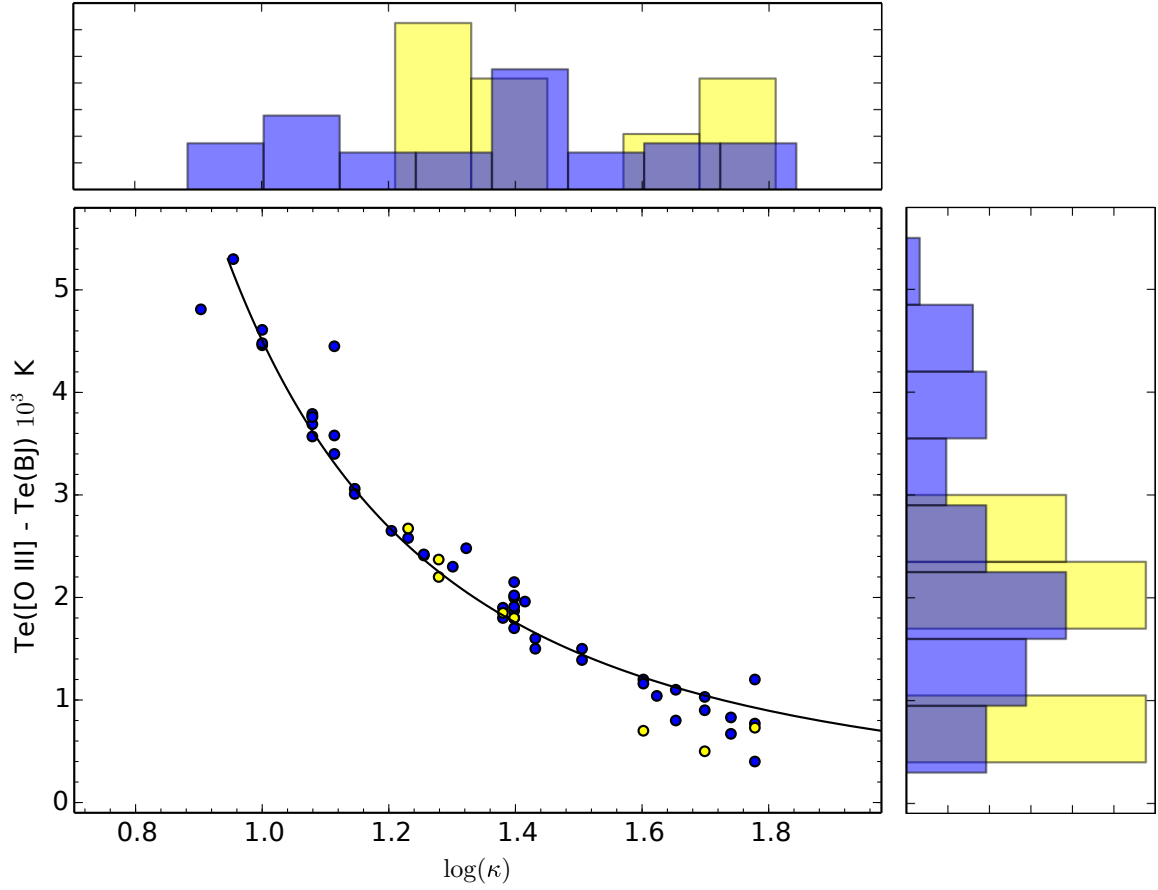}
\caption{The electron temperature discrepancies, $T_{\rm e}$([\ion{O}{3}])$-T_{\rm e}$({\rm BJ}),
are plotted against the $\kappa$ values for PNe  (blue) and \ion{H}{2} regions (yellow).
The histograms in the top and right panels show their distributions.
\label{fig2}}
\end{figure}

Tables~\ref{PN} \& \ref{H2} also list $T_{\rm e}({\rm BJ})$ and $T_{\rm e}([{\rm O~III}])$
reported in the literature. As quoted, $T_{\rm e}({\rm BJ})$ and $T_{\rm e}([{\rm O~III}])$ are
respectively determined from $J$(BJ) and $R$([\ion{O}{3}]) under the assumption of MB EEDs.
In the scenario of $\kappa$ EEDs, the two parameters retrieved from
$J$(BJ) and $R$([\ion{O}{3}]) are substituted with $\kappa$ and $T_{\rm U}$. Namely, a consistent
temperature $T_{\rm U}$ can be  obtained through adjusting the $\kappa$ index. 
Therefore, $T_{\rm U}$ should be a value between $T_{\rm e}({\rm BJ})$ and $T_{\rm e}([{\rm O~III}])$,
as confirmed in those tables. Since the $\kappa$ index is introduced for the purpose of interpreting
 the temperature discrepancy, $\delta t=T_{\rm e}([{\rm O~III}])-T_{\rm e}({\rm BJ})$,
it is expected to be negatively correlated with $\delta t$. Such a correlation can be 
seen in Figure~\ref{fig2}. The $\kappa$ index can be fitted to the expression
\begin{equation}\label{fitkappa}
\kappa = 1.5\exp\left(\frac{3.57}{\delta t^{0.42}}\right),
\end{equation}
where $\delta t$ is in the unit of $10^{-3}$\,K. Equation~(\ref{fitkappa}) can be used to 
conveniently determine the $\kappa$ index from the previously reported $T_{\rm e}({\rm BJ})$ 
and $T_{\rm e}([{\rm O~III}])$. Figure~\ref{fig2} also shows the histograms of $\kappa$ 
and $\delta t$, where PNe clearly exhibit an extended tail towards low $\kappa$ and high
$\delta t$, and there is no \ion{H}{2} region showing $\kappa<17$.

\begin{figure}
\plotone{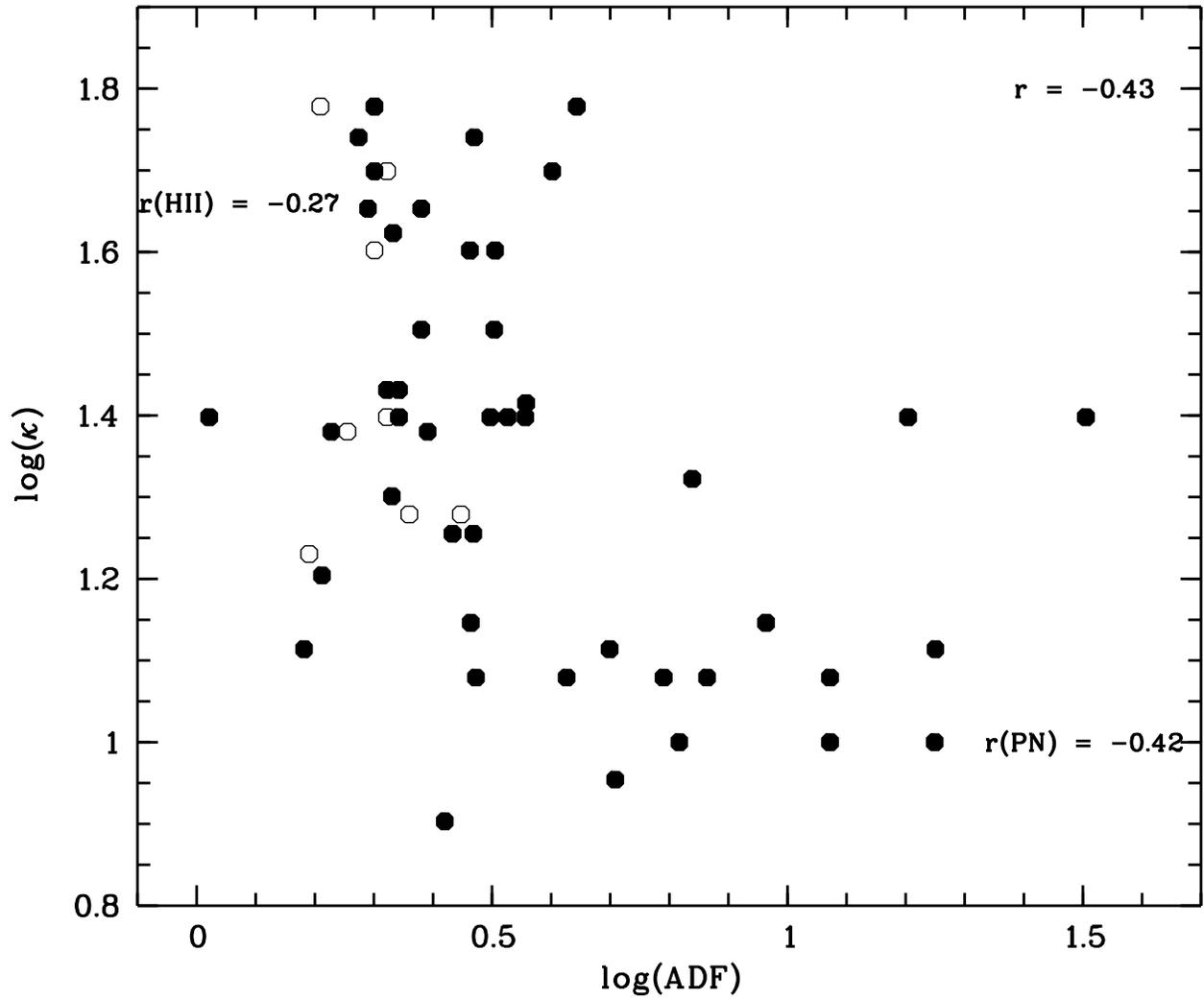}
\caption{$\log\kappa$ versus $\log({\rm ADF})$ for PNe (filled circles) and
\ion{H}{2} regions (open circles) with corresponding correlation coefficients
 indicated. 
The correlation coefficient for the combined data is given in the
upper right corner.
\label{kvadf}}
\end{figure}

It has been well established that the RL/CEL temperature and abundance discrepancies 
are two relevant problems.  A positive  correlation between ADFs and $\delta t$  has 
been found in a small sample of PNe by \citet{ll01}, suggesting that the two discrepancies are
probably caused by a common underlying physical mechanism. If this is the case,
 we would expect that there exists a negative correlation between $\kappa$ and ADFs. 
Figure~\ref{kvadf} plots $\kappa$ against ADFs, in which we indeed observe a
loose negative correlation for our PN sample.  The relationship between 
$\log(\kappa)$ and $\log$(ADF) is clearly non-linear, but similar to that
between $\log(\kappa)$ and $\delta t$ (Figure~\ref{fig2}), shows a `L'-shape in Figure~\ref{kvadf}.
This is in agreement with \citet{ll01} who found a strong linear correlation
between $\delta t$ and $\log$(ADF).
For \ion{H}{2} regions, however, the correlation coefficient is too low to be meaningful.

\section{DISCUSSION}

The $\kappa$ distribution provides a potential solution for the RL/CEL discrepancy 
problem, as illustrated in Figure~\ref{fig1}.  
 Using Equation~(\ref{eq3}), our results show
that about $3.2\%$ and $2.7\%$ electrons are non-thermal in PNe and \ion{H}{2} regions,
respectively, suggesting that only minority of non-thermal electrons are able to
account for the $T_{\rm e}$(BJ)/$T_{\rm e}$([\ion{O}{3}]) discrepancy.
\citet{nic12,nic13} discussed the possible
formation of $\kappa$  EEDs in gaseous nebulae.
Because the thermalization timescale of free 
electrons is proportional to the cube of the velocity, high-energy electrons may approach to
equilibration state slower than their injection.  Therefore, if energetic electrons could
be continually and quickly injected, the $\kappa$ distribution would be developed.
A key question is what the mechanism continually pumping and maintaining stable $\kappa$ EEDs 
could be in PNe and \ion{H}{2} regions.  The possibilities presented by \citet{nic12,nic13}
include magnetic reconnections, local shocks, photoionization of dust, and X-ray ionization.
The $\kappa$ indices determined in the present paper provide a useful foundation for 
investigating the postulated generation of non-thermal electrons. For that purpose, in this section
we  examine the correlations between $\kappa$ indices and other nebular properties.

\begin{figure}
\plotone{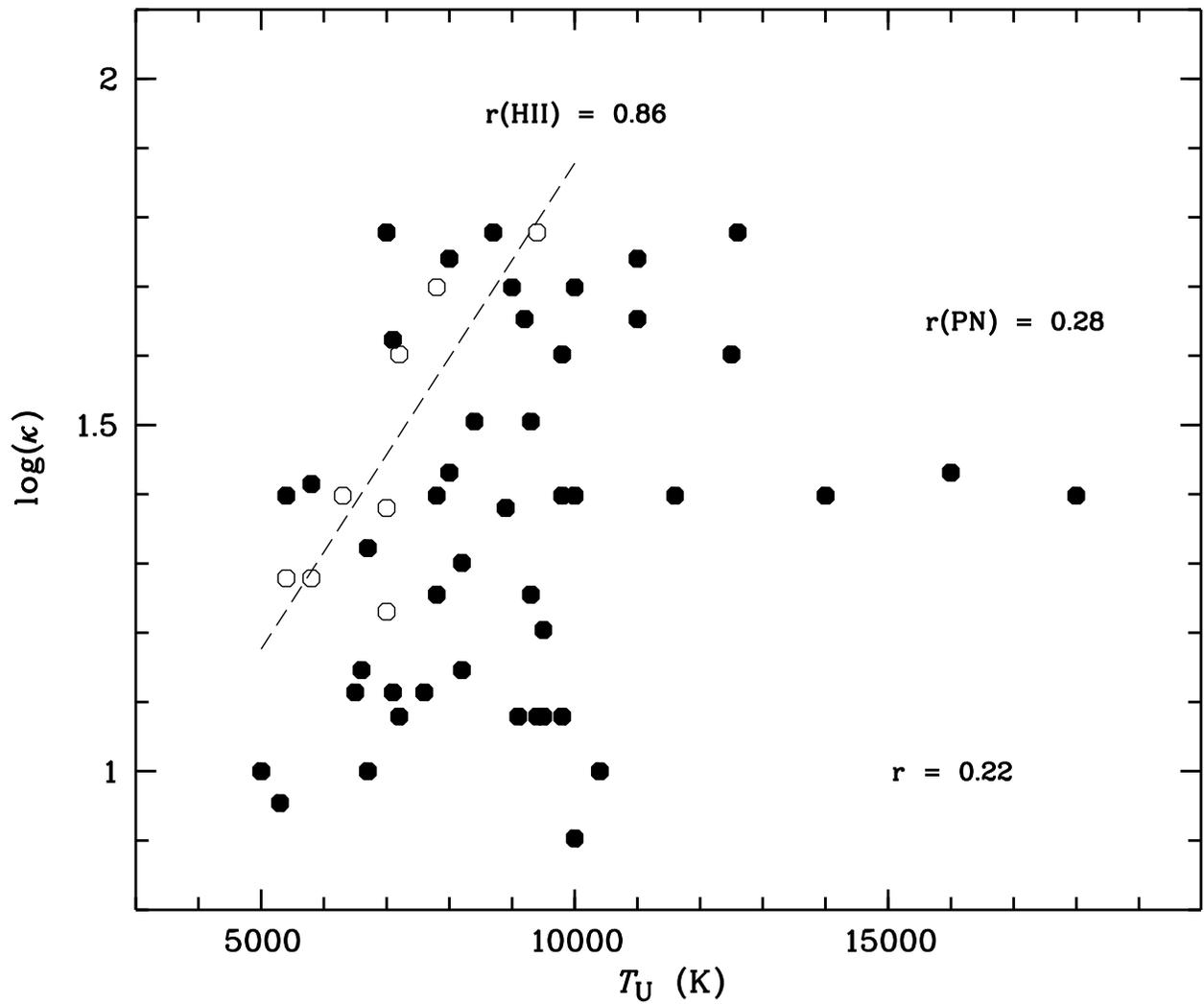}
\caption{$\log\kappa$ versus $T_{\rm U}$ for PNe (filled circles) and
\ion{H}{2} regions (open circles). The dashed line is
the best linear fit  for \ion{H}{2} regions. The correlation coefficients for
PNe, \ion{H}{2} regions, and the combined data  are given.
\label{kvt}}
\end{figure}

Figure~\ref{kvt} shows that there is no apparent linear correlation between $\kappa$ and 
$T_{\rm U}$ for PNe. 
For cooler nebulae, a more significant fraction of electrons lying in the power-law 
high-energy tail contributes to the excitation of [\ion{O}{3}] CELs in that the MB core is confined within 
a lower-energy region and thus has less contribution to the collisional excitation. 
This points to a positive correlation between $\kappa$ and $T_{\rm U}$.
However, the situation is complicated by the fact that the suprathermal heating  caused by energetic stellar 
winds can increase the kinetic temperature and  decrease the $\kappa$ index,
as suggested by \citet{nic12}.  A visual examination of Figure~\ref{kvt} seems to 
 reveal that the $\kappa$ index roughly increases with increasing $T_{\rm U}$, but
significantly decreases at the highest temperature. Although this behavior can be
explained in terms of  $\kappa$ EEDs, it cannot be viewed as a solid support for 
the $\kappa$ distribution. Figure~\ref{kvt} also suggests 
a stronger positive correlation for \ion{H}{2} regions than PNe. A natural question to ask is whether
the cause of RL/CEL discrepancies in \ion{H}{2} regions differs from that in
PNe.  We require a larger sample  and more sophisticated study to ascertain this.

\begin{figure}
\plotone{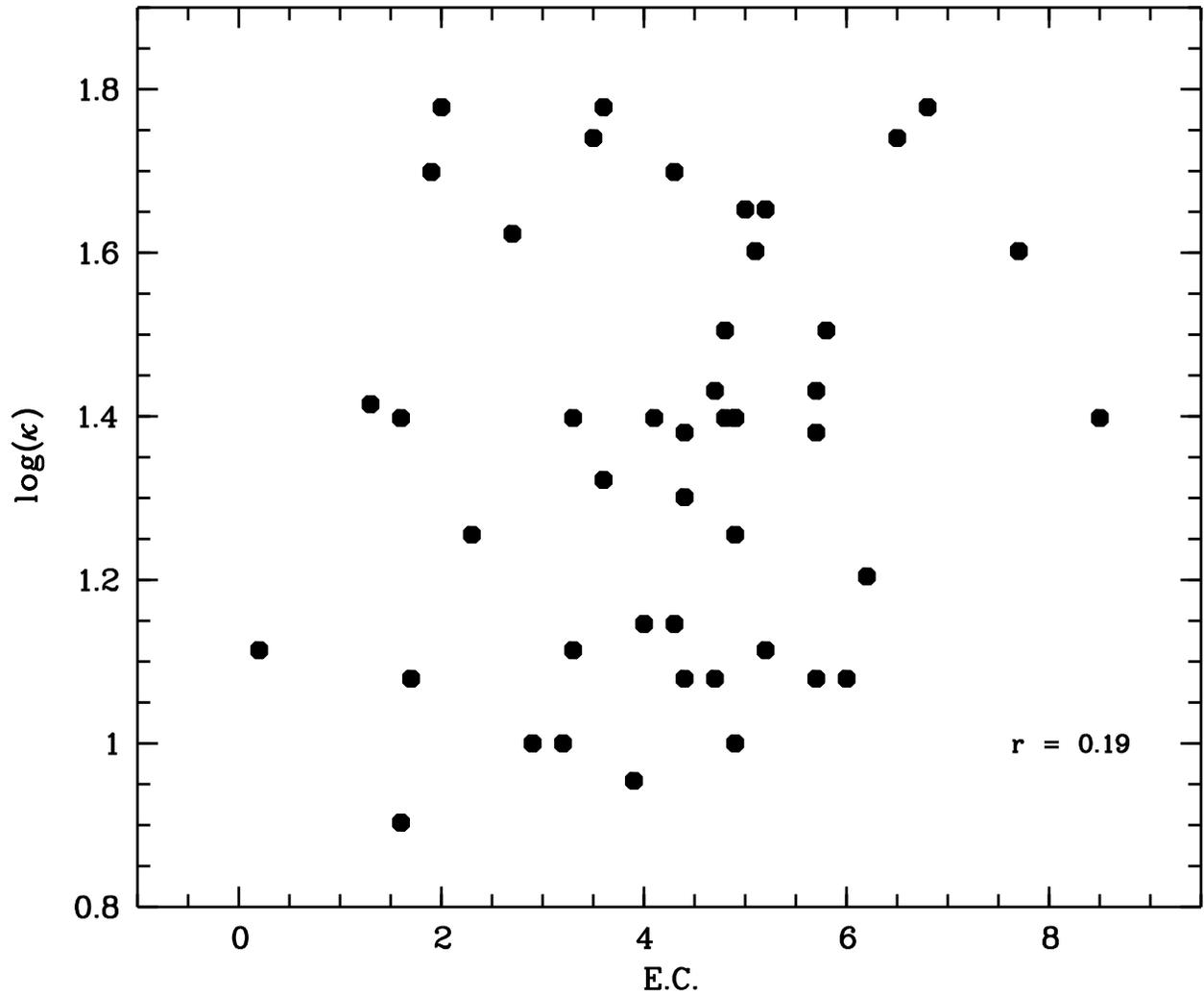}
\caption{$\log\kappa$ versus E.C. for PNe.
The correlation coefficient is indicated.
\label{kvec}}
\end{figure}

In Figure~\ref{kvec} we plot $\kappa$ against the excitation class (EC) of PNe.
The excitation class can be determined from spectral line ratios, and is closely
related to the effective temperature of the central stars. We calculated the ECs of PNe
 following the formalism suggested by \citet{dopita}, as tabulated in 
Table~\ref{PN}. Although the classification scheme  was developed 
for the Magellanic Cloud PNe,  it should be able to serve as a measurement of 
the relative excitation conditions in Galactic PNe. Figure~\ref{kvec} demonstrates
that no correlation exists between $\kappa$  and EC. We therefore do not find a trend that
more  pronounced departure from MB EEDs corresponds to harder radiation fields.
Consequently, we cannot give any empirical evidence for the photoionzation
by the radiation from central stars as the cause of  non-thermal electron production.

Is it possible that the non-thermal electrons are generated by the kinetic energy
released by the mass loss?  NGC\,40 and NGC\,1501 are two PNe with Wolf-Rayet type 
central stars that are characterized by fast stellar winds and high mass-loss rates.
We only discover moderate  $\kappa$ indices for the two PNe, while those with the lowest 
$\kappa$ indices are not Wolf-Rayet PNe. Furthermore, \citet{gp13} investigated the
abundances of a sample of PNe with [WC]-type nucleus, and found that there is no discernible 
relation between the [WC] nature and the ADFs. Therefore, we can conclusively rule out the
possibility that stellar winds are the main source producing non-thermal electrons.

We also examine the $\kappa$ indices by classifying the PNe according to their morphologies
and Peimbert types (see the 10th column of Table~\ref{PN}). The morphologies of our sample PNe 
can be classified as bipolar (B), elliptical (E), round (R), and quasi-stellar (S).
Their mean $\kappa$ indices are  $29\pm14$, $24\pm15$, $35\pm16$, and $19\pm20$, respectively. Given the large standard deviations and 
small sample numbers in each group, we do not detect statistically significant differences 
of the $\kappa$ indices between the PNe of different morphologies. 
The Peimbert type
\citep{pei78} can roughly reflect the stellar population of the Galaxy. We used the
Peimbert-type classification method introduced by \citet{qr07}. Type I PNe descend 
from high-mass progenitors, and represent the youngest population, while type IV PNe
represent the oldest population in halo. We derive the mean $\kappa$ index of type I PNe to 
be $34\pm18$, slightly larger than that of non-type I PNe ($26\pm15$). 
There are a few type IV PNe exhibiting very low $\kappa$ indices ($<10$).
This is consistent with previous findings that young PNe have systematically 
smaller ADFs \citep[e.g.,][]{zl05}. If the $\kappa$ distribution holds in PNe, it is 
hard to understand why more deviations from thermal equilibrium can be developed in older 
PNe.

\begin{figure*}
\begin{center}
\epsfig{file=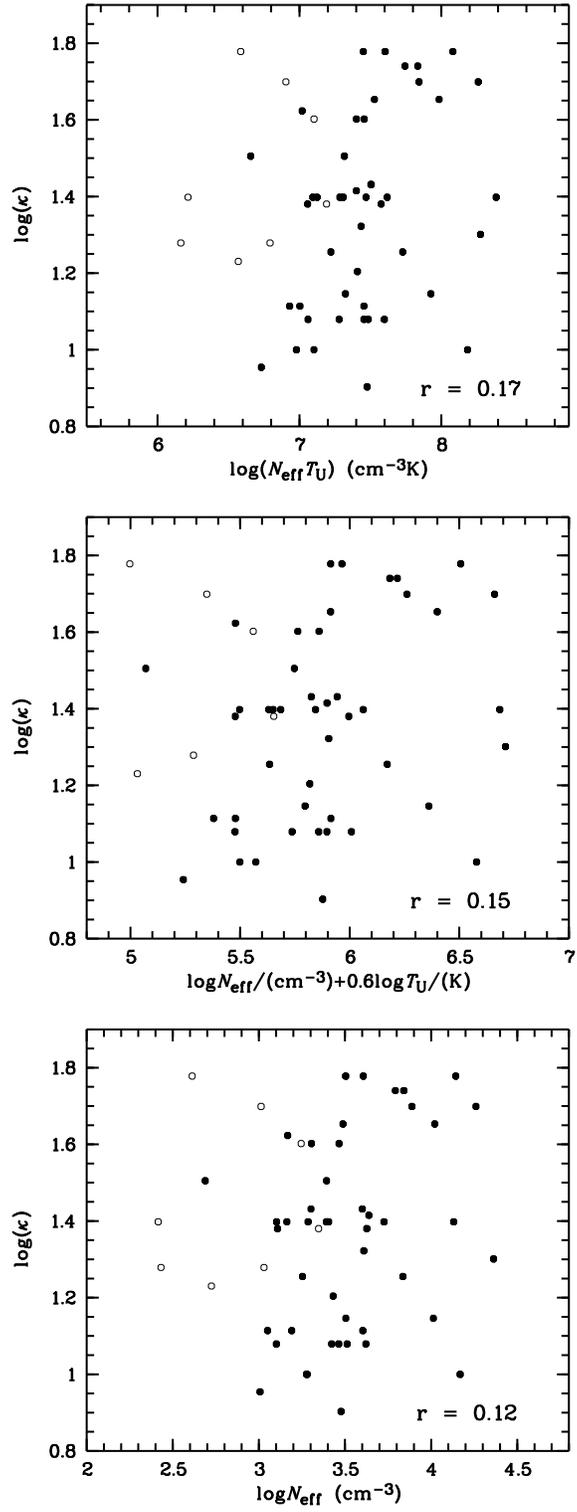,
height=20cm}
\caption{$\log \kappa$ versus $\log (N_{\rm eff}T_{\rm U}^\nu)$
for PNe (filled circles) and \ion{H}{2} regions (open circles), where
$\nu=1$ (upper), 0.6 (middle), and 0(lower). 
The correlation coefficients are given in the lower right corner of each 
panel.
\label{kvnt} }
\end{center}
\end{figure*}

To reside in stationary states out of thermal equilibrium, the energetic electrons must be 
non-collisional. Collisionless plasma can be characterized by small ratio between the
Debye length, $\lambda_{\rm D}\sim(T_{\rm U}/N_{\rm eff})^{0.5}$ and the mean free path,
$L_{\rm m}\sim T_{\rm U}^2/N_{\rm eff}$, i.e., 
$\lambda_{\rm D}/L_{\rm m}$ ($\sim N_{\rm eff}^{0.5}/T_{\rm U}^{1.5}$) lower than one. Therefore,
the $\kappa$ index should be related to the density and temperature of the system.
\citet{liv15} found a negative correlation between  $M$ [here, $M=1/(\kappa-0.5)$] and $\log (N_{\rm eff}T_{\rm U}^\nu)$
for $\nu=1$, 0.6, and 0 (see their Figure~2).  The sample examined by \citet{liv15} includes various space 
plasmas, but most of them are solar system plasmas.
In Figure~\ref{kvnt}, we examine the correlation between $\log \kappa$ and $\log (N_{\rm eff}T_{\rm U}^\nu)$
for our nebula sample. Although these best linear fits seem to show a trend that $\log \kappa$ increases 
with increasing $\log (N_{\rm eff}T_{\rm U}^\nu)$, the point distributions in these diagrams are too scattering
to definitely indicate a correlation. When comparing with Figure~2 of \citet{liv15}, our data mostly concentrate
in the right-down corner, namely the regions centred at $(\log N_{\rm eff} {[\rm m^{-3}]}, M)=(9.5, 0.05)$,
$\log (N_{\rm eff}[\rm m^{-3}]T_{\rm U}^{0.6}[\rm K], M)=(11.8, 0.05)$, and 
$\log (N_{\rm eff}[\rm m^{-3}]T_{\rm U}[\rm K], M)=(13.5, 0.05)$. The $\kappa$ indices obtained in nebule are
generally larger than those in solar system plasma.  It should be noted that our plots are not contrary to those
of \citet{liv15} as they examined a much wider parameter space. However, the non-existent correlation
between the $\kappa$ index and the pair $(N_{\rm eff}, T_{\rm U})$ in a smaller parameter space
casts some doubts on whether $\kappa$ EEDs hold in PNe and \ion{H}{2} regions.

ZLZ determined the $\kappa$ indices of four PNe through fitting their \ion{H}{1} FB continuities. Three of them
are included in our sample. However, the present method yields larger $\kappa$ indices for the three PNe
than those obtained by ZLZ.
Because the \ion{H}{1} FB emission samples the electrons with lower energy than the [\ion{O}{3}] CELs do,
if the physical conditions of PNe are inhomogeneous we may obtained different results  from the two methods.
Therefore, it seems inappropriate to use a single $\kappa$ value to characterize the EED of a given PN.
To further clarify this point, we need to investigate the behavior of other temperature diagnostic lines in
$\kappa$ EEDs, such as the [\ion{N}{2}] CEL ratio. The main difficulty to perform such a study is that
the cross sections of collision are difficult to be tabulated, and thus are usually unavailable in the literature.
Recently, \citet{hs15} presented a method to approximate $\kappa$ distributions as a sum of MB distributions, 
which provides an easy way to convert the existing rate coefficients with MB EEDs to those with $\kappa$ EEDs.
The constraint of other CELs on the the $\kappa$ indices will be the subject of a forthcoming paper.

\section{SUMMARY}

Assuming that the discrepancy between $T_{\rm e}$(BJ) and $T_{\rm e}$([\ion{O}{3}]) is completely 
attributed to non-thermal EEDs, we determine the $\kappa$  indices for a sample of PNe and \ion{H}{2}
regions. This is for the first time that the $\kappa$ indices for a large nebula sample are reported.
These data provide a valuable resource for further research of the non-thermal electron distribution in
space plasmas.  Our results show that PNe have systematically lower  $\kappa$  indices than \ion{H}{2} 
regions, and the $\kappa$  indices in nebulae are significantly larger than those in solar system 
plasmas. Through an empirical fitting, we also present a convenient formula to deduce the $\kappa$  index
from the $T_{\rm e}$(BJ)/$T_{\rm e}$([\ion{O}{3}]) discrepancy.

Although the $\kappa$ EED provides a promising way to explain the long-standing RL/CEL discrepancy 
problem, its origin and  physics validity should be thoroughly investigated.  In order to explore
the possible mechanisms that can cause the formation of the $\kappa$ distribution in photoionized
gaseous nebulae, we examine the correlation between the obtained  $\kappa$  indices and various other 
nebular properties. However, we cannot find sound evidence supporting that non-thermal electrons can be 
pumped in PNe and \ion{H}{2} regions. For three extreme PNe, the currently obtained
$\kappa$ indices are larger than those by ZLZ utilizing the \ion{H}{1} FB continuum.
In order to interpret this discrepancy within the framework of  $\kappa$ EEDs,
spatial variations of $\kappa$ and $T_{\rm U}$ are required.
Given the scale of $\kappa$ distributions in the solar system, it is possible that such distributions may be present over small scales in nebular regions,
and would be likely detected, should they exist, in nearby nebulae
utilizing future high-resolution high-sensitivity facilities such as the James Webb Space Telescope (JWST).

Despite the difficulties to identify the orgin of non-thermal electrons, the present study
cannot rule out the existence of the $\kappa$ distribution in PNe and \ion{H}{2} regions. This scenario
provides an intriguing possibility to solve some observational puzzles in nebulae.
The $\kappa$ distribution can greatly affect thermal and ionization structures of nebulae and, if proven true, should be incorporated into
photoionization models. One such attempt has been made by
\citet{ds13} who modified the photoionization code, 
MAPPINGS,  to investigate the effect of $\kappa$ EEDs on abundance 
determination of \ion{H}{2} regions. It would be a useful addition to 
applications such as Cloudy \citep{fp13}, as a means of exploring the possible 
diagnostic symptoms of  $\kappa$ EEDs.
Apparently, it is extremely important to develop methods to detect EEDs from 
observations.  
We hope that the results reported in this paper can serve as a useful
reference to further address this issue.

\acknowledgments

We thank the anonymous referee for a positive review of the manuscript,
and Dr. Bojicic Ivan for useful discussion on the classification of PNe.
Financial support for this work was provided by the Research Grants 
Council of the Hong Kong under grants HKU7073/11P and HKU7062/13P.

\clearpage

\clearpage

\clearpage



\clearpage

\begin{deluxetable}{lccccccccccr}
\tabletypesize{\scriptsize}
\tablewidth{0pt}
\tablecaption{The derived $\kappa$ values and other properties of PNe.\label{PN}}
\tablehead{
\colhead{Object}  & \colhead{T$_{\rm e}$({\rm BJ})} &
\colhead{T$_{\rm e}$([{\rm O}~III])}  &
\colhead{N$_{\rm e}$} &
\colhead{$\kappa$} &
\colhead{T$_{\rm U}$}  & 
\colhead{N$_{\rm eff}$} &
\colhead{E.C.}  &
\colhead{ADF} & \colhead{Class$^a$} &
\colhead{Ref.}\\
 & (K) & (K) &(cm$^{-3}$)  & & (K) & (cm$^{-3}$)}
\startdata
BoBn 1    &    8840   &   13650 &3370  &    8   &  10000  &3000 &   1.6    &   2.63   &S,{\sc iv}  & O10  \\
Cn 1-5    &   10000   &    8770 &3391  &\nodata & \nodata &3391 &   3.9    &   1.02   &B,{\sc ii}  & W07 \\
Cn 2-1    &   10800   &   10250 &10315 &\nodata & \nodata &10315&  \nodata &   1.02   &S,{\sc iii} & W07 \\
Cn 3-1    &    5090   &    7670 &6830  &   17   &   5500  &6494 &  \nodata & \nodata  &E,{\sc ii}  & W05   \\
DdDm 1    &    8730   &   12300 &4500  &   12   &   9500  &4180 &   1.7    &  11.80   &E,{\sc iv}  & W05 \\
H 1-35    &   12000   &    9060 &22585 &\nodata & \nodata &22585&   2.3    &   1.04   &S,{\sc iv}  & W07 \\
H 1-41    &    4500   &    9800 &1125  &    9   &   5300  &1016 &   3.9    &   5.11   &S,{\sc iii} & W07 \\
H 1-42    &   10000   &    9690 &7508  &$>$60   & \nodata &7508 &   4.2    &   1.04   &B,{\sc ii}  & W07 \\
H 1-50    &   12000   &   10950 &9355  &\nodata & \nodata &9355 &   5.0    &   1.05   &E,{\sc ii}  & W07 \\
H 1-54    &   12500   &    9540 &13032 &\nodata & \nodata &13032&   2.0    &   1.05   &B,{\sc iv}  & W07 \\
He 2-118  &   14500   &   12630 &15155 &\nodata & \nodata &15155&  \nodata &   1.03   &S,{\sc iii} & W07 \\
Hu 1-1    &    8350   &   12110 &1360  &   12   &   9100  &1263 &   4.7    &   2.97   &E,{\sc ii}  & W05  \\
Hu 1-2    &   18900   &   19500 &4467  &\nodata & \nodata &4467 &   3.4    &   1.60   &B,{\sc ii}  & L04  \\
Hu 2-1    &    8960   &    9860 &7870  &   50   &   9000  &7742 &   1.9    &   4.00   &B,{\sc ii}  & W05  \\
IC 351    &   11050   &   13070 &2630  &   25   &  11600  &2543 &   4.9    &   3.14   &E,{\sc ii}  & W05  \\
IC 1747   &    9650   &   10850 &2980  &   40   &   9800  &2919 &   5.1    &   3.20   &E,{\sc ii}  & W05 \\
IC 2003   &    8960   &   12650 &3130  &   12   &   9800  &2908 &   4.4    &   7.31   &E,{\sc ii}  & W05  \\
IC 3568   &    9490   &   11400 &1995  &   25   &  10000  &1929 &   4.8    &   2.20   &R,{\sc ii}  & L04 \\
IC 4191   &    9200   &   10000 &10695 &   45   &   9200  &10501&   5.0    &   2.40   &B,{\sc ii}  & T03 \\
IC 4406   &    9350   &   10000 &1560  &$>$60   & \nodata &1560 &   4.7    &   1.90   &B,{\sc ii}  & T03 \\
IC 4699   &   12000   &   11720 &2119  &\nodata & \nodata &2119 &   5.5    &   1.09   &E,\nodata   & W07 \\
IC 4846   &    7700   &   10710 &10960 &   14   &   8200  &10299&   4.3    &   2.91   &B,{\sc iii} & W05  \\
IC 5217   &   11350   &   11270 &4510  &\nodata & \nodata &4510 &   \nodata&   2.26   &E,{\sc iii} & W05  \\
M 1-20    &   12000   &    9860 &10151 &\nodata & \nodata &10151&   4.3    &   1.02   &E,{\sc ii}  & W07 \\
M 1-29    &   10000   &   10830 &6297  &   55   &  11000  &6204 &   6.5    &   2.95   &E,{\sc ii}  & W07 \\
M 1-61    &    9500   &    8900 &20817 &\nodata & \nodata &20817&   4.0    &   1.03   &R,{\sc ii}  & W07 \\
M 1-73    &    5490   &    7450 &4490  &   26   &   5800  &4347 &   1.3    &   3.61   &B,\nodata   & W05  \\
M 1-74    &    7850   &   10150 &24030 &   20   &   8200  &23032&   4.4    &   2.14   &R,{\sc ii}  & W05  \\
M 2-4     &    7900   &    8570 &7041  &   55   &   8000  &6937 &   3.5    &   1.88   &S,{\sc ii}  & W07 \\
M 2-6     &   11700   &   10100 &7523  &\nodata & \nodata &7523 &   3.1    &   1.04   &E,{\sc ii}  & W07  \\
M 2-27    &   14000   &   11980 &11217 &\nodata & \nodata &11217&   4.2    &   1.04   &E,{\sc iii} & W07 \\
M 2-31    &   14000   &    9840 &6141  &\nodata & \nodata &6141 &   \nodata& \nodata  &R,\nodata   & W07 \\
M 2-33    &    7000   &    8040 &1501  &   42   &   7100  &1471 &   2.7    &   2.150  &E,{\sc iv}  & W07 \\
M 2-36    &    5900   &    8380 &4230  &   21   &   6700  &4063 &   3.6    &   6.90   &B,{\sc ii}  & L01  \\
M 2-42    &   14000   &    8470 &3430  &\nodata & \nodata &3429 &   3.6    &   1.04   &E,{\sc iii} & W07 \\
M 3-7     &    6900   &    7670 &4093  &   60   &   7000  &4037 &   2.0    &   4.40   &R,{\sc iv}  & W07 \\
M 3-21    &   10400   &    9790 &13652 &\nodata & \nodata &13652&   \nodata&   1.05   &E,{\sc ii}  & W07 \\
M 3-29    &   10700   &    9190 &813   &\nodata & \nodata &813  &   2.5    &   1.04   &E,{\sc ii}  & W07 \\
M 3-32    &    4400   &    8860 &2085  &   10   &   5000  &1905 &   3.2    &  17.75   &E,{\sc iv}  & W07 \\
M 3-33    &    5900   &   10380 &2068  &   10   &   6700  &1889 &   4.9    &   6.56   &E,{\sc iii} & W07  \\
M 3-34    &    8440   &   12230 &3500  &   12   &   9400  &3251 &   5.7    &   4.23   &S,\nodata   & W05  \\
Me 2-2    &   10590   &   10970 &11930 &$>$60   & \nodata &11930&   \nodata&   2.10   &B,{\sc ii}  & W05  \\
NGC 40    &    7020   &   10600 &1202  &   13   &   7600  &1123 &   0.2    &  17.80   &E,{\sc ii}  & L04 \\
NGC 1501  &    9400   &   11100 &1312  &   25   &   9800  &1268 &   4.9    &  32.00   &E,{\sc i}   & E04  \\
NGC 2022  &   13200   &   15000 &1505  &   25   &  14000  &1455 &   3.3    &  16.00   &B,{\sc ii}  & T03  \\
NGC 2440  &   14000   &   16150 &\nodata&  25   &  15000  &\nodata& 8.3    &   5.40   &E,{\sc ii}  & T03   \\
NGC 2867  &    8950   &   11600 &2850  &   16   &   9500  &2700 &   6.2    &   1.63   &B, {\sc ii} & G09 \\
NGC 3132  &   10780   &    9530 &600   &\nodata & \nodata &600  &   3.6    &   3.50   &B,{\sc ii}  & T03  \\
NGC 3242  &   10200   &   11700 &2070  &   27   &  16000  &2006 &   5.7    &   2.20   &B,{\sc ii}  & T03  \\
NGC 3918  &   12300   &   12600 &5667  &$>$60   & \nodata &5667 &   6.6    &   2.30   &B,{\sc ii}  & T03  \\
NGC 5307  &   10700   &   11800 &3133  &   45   &  11000  &3076 &   5.2    &   1.95   &B,{\sc i}   & R03  \\
NGC 5315  &    8600   &    9000 &14091 &   60   &   8700  &13900&   3.6    &   2.00   &B,{\sc i}   & T03  \\
NGC 5882  &    7800   &    9400 &4113  &   27   &   8000  &3987 &   4.7    &   2.10   &B,{\sc ii}  & T03  \\
NGC 6153  &    6080   &    9140 &3400  &   14   &   6600  &3195 &   4.0    &   9.20   &E,{\sc i}   & L00  \\
NGC 6210  &    9300   &    9680 &4365  &$>$60   & \nodata &4365 &   3.5    &   3.10   &B,{\sc iii} & L04  \\
NGC 6302  &   16400   &   18400 &14000 &   25   &  18000  &13538&   8.5    &   3.60   &B,{\sc i}   & T03  \\
NGC 6439  &    9900   &   10360 &5169  &$>$60   & \nodata &5169 &   5.7    &   6.16   &E,{\sc iii} & W07 \\
NGC 6543  &    8340   &    8000 &4770  &\nodata & \nodata &4770 &   3.1    &   4.20   &B,{\sc i}   & W04   \\
NGC 6565  &    8500   &   10300 &1329  &   24   &   8900  &1283 &   5.7    &   1.69   &E,{\sc ii}  & W07 \\
NGC 6567  &   14000   &   10580 &8118  &\nodata & \nodata &8118 &   4.2    &   1.04   &E,{\sc iii} & W07 \\
NGC 6572  &   11000   &   10600 &15136 &\nodata & \nodata &15136&   \nodata&   1.60   &B,{\sc ii}  & L04  \\
NGC 6620  &    8200   &    9590 &2535  &   32   &   8400  &2470 &   5.8    &   3.19   &R,{\sc ii}  & W07 \\
NGC 6720  &    9100   &   10600 &501   &   32   &   9300  &488  &   4.8    &   2.40   &E,{\sc ii}  & L04  \\
NGC 6741  &   15300   &   12600 &5129  &\nodata & \nodata &5129 &   6.3    &   1.90   &E,{\sc ii}  & L04  \\
NGC 6790  &   15000   &   12800 &39811 &\nodata & \nodata &39811&   \nodata&   1.70   &B,{\sc ii}  & L04  \\
NGC 6803  &    7320   &    9740 &7190  &   18   &   7800  &6857 &   4.9    &   2.71   &E,{\sc ii}  & W05  \\
NGC 6807  &    9900   &   10930 &18530 &   50   &  10000  &18229&   4.3    &   2.00   &R,{\sc iv}  & W05  \\
NGC 6818  &   12140   &   13300 &2063  &   40   &  12500  &2020 &   7.7    &   2.90   &E,{\sc ii}  & T03  \\
NGC 6826  &    9650   &    9370 &1995  &\nodata & \nodata &1995 &   3.4    &   1.90   &E,{\sc ii}  & L04 \\
NGC 6833  &   13670   &   12810 &19030 &\nodata & \nodata &19030&  \nodata &   2.47   &B,{\sc iv}  & W05  \\
NGC 6879  &    8500   &   10400 &4380  &   24   &   8900  &4229 &   4.4    &   2.46   &R,{\sc ii}  & W05  \\
NGC 6884  &   11600   &   11000 &7413  &\nodata & \nodata &7412 &   5.3    &   2.30   &E,{\sc ii}  & L04 \\
NGC 6891  &    5930   &    9330 &1660  &   13   &   6500  &1551 &   3.3    &   1.52   &E,{\sc ii}  & W05  \\
NGC 7009  &    6490   &   10940 &4290  &   13   &   7100  &4010 &   5.2    &   5.00   &B,{\sc ii}  & F11  \\
NGC 7026  &    7440   &    9310 &5510  &   25   &   7800  &5328 &   4.1    &   3.36   &B,{\sc ii}  & W05  \\
NGC 7027  &   12800   &   12600 &52289 &$>$60   &  \nodata&51589&   7/0    &   1.29   &B,{\sc ii}  & Z05   \\    
NGC 7662  &   12200   &   13400 &3236  &   60   &  12600  &3192 &   6.8    &   2.00   &E,{\sc ii}  & L04 \\
PB 8      &    5100   &    6900 &2550  &   25   &   5400  &2465 &   1.6    &   1.05   &E,{\sc ii}  & G09 \\
Sp 4-1    &    8830   &   11240 &1880  &   18   &   9300  &1792 &   2.3    &   2.94   &B,\nodata   & W05  \\
Vy 1-2    &    6630   &   10400 &2850  &   12   &   7200  &2647 &   6.0    &   6.17   &E,{\sc iv}  & W05  \\
Vy 2-1    &    8700   &    7860 &3815  &\nodata & \nodata &3815 &   2.8    &   1.03   &E,{\sc ii}  & W07  \\
Vy 2-2    &    9300   &   13910 &16130 &   10   &  10400  &14740&   2.9    &  11.80   &S,{\sc iv}  & W05  \\
\enddata
\tablenotetext{a}{Morphology classes (B: bipolar; E: elliptical; R: round; S: quasi-stellar), 
Peimbert types ({\sc i}--{\sc iv}).}
\tablerefs{
(E04) Ercolano et al. 2004; (F11) Fang \& Liu 2011; 
(G09) Garc\'ia-Rojas et al. 2009;
(L00) Liu et al. 2000;
(L01) Liu et al. 2001; (L04) Liu et al. 2004; (O10) Otsuka et al. 2010;
(R03) Ruiz et al. 2003; (T03) Tsamis et al. 2003a; 
(W07) Wang \& Liu 2007; 
(W04) Wesson \& Liu 2004;
(W05) Wesson et al. 2005;
(Z05) Zhang et al. 2005.}
\end{deluxetable}

\begin{deluxetable}{lcccccccr}
\tabletypesize{\scriptsize}
\tablewidth{0pt}
\tablecaption{The derived $\kappa$ values and other properties of \ion{H}{2} regions.\label{H2}}
\tablehead{
\colhead{Object}  & 
\colhead{T$_{\rm e}$({\rm BJ})} &
\colhead{T$_{\rm e}$([{\rm O~III}])}  &
\colhead{N$_{\rm e}$} &
\colhead{$\kappa$} &
\colhead{T$_{\rm U}$}  &
\colhead{N$_{\rm eff}$} &
\colhead{ADF} & 
\colhead{Ref.}\\
 & (K) & (K) &(cm$^{-3}$)  & & (K) & (cm$^{-3}$)}
\startdata
30 Dor   &  9220 &        9950   & 416  &        60  &     9400  &  410  &   1.62   & P03   \\
H 1013   &  5000 &        7700   & 280  &        19  &     5400  &  270  &   2.29   & B07,E09 \\
M 8      &  7100 &        7800   & 1800 &        40  &     7200  &  1760 &   2.0    & G05 \\
M 16     &  5450 &        7650   & 1120 &        19  &     5800  &  1070 &   2.8    & G07 \\
M 17     &  7700 &        8200   & 1050 &        50  &     7800  &  1030 &   2.1    & T03 \\
M 20     &  6000 &        7980   & 270  &        25  &     6300  &  260  &   2.1    & G07 \\
M 42     &  7900 &        8300   & 6350 &      $>60$ &     8000  &  6260 &   1.02   & E04 \\
NGC 3576 &  6650 &        8500   & 2300 &        24  &     7000  &  2220 &   1.8    & G07 \\
NGC 5447 &  6610 &        9280   & 560  &        17  &     7000  &  530  &   1.55   & E09 \\
S 311    &  9500 &        9000   & 310  &       ...  &     ...   &  310  &   1.03   & G05 \\
\enddata
\tablerefs{
(B07) Bresolin 2007; (E04) Esteban et al. 2004; (E09) Esteban et al. 2009; (G05) 
Garc\'ia-Rojas et al. 2005; (G07) Garc\'ia-Rojas \& Esteban 2007; (P03) Peimbert 2003; (T03) Tsamis et al. 2003b.
}
\end{deluxetable}

\end{document}